\documentclass[12pt]{article}
\usepackage{latexsym}
\newcommand{\Sin}{\sin}
\newcommand{\Cos}{\cos}

\begin{document}
\input epsf
\def\be{\begin{equation}}
\def\bea{\begin{eqnarray}}
\def\ee{\end{equation}}
\def\eea{\end{eqnarray}}
\def\d{\partial}
\def\la{\lambda}
\def\eps{\epsilon}
\def\a{{\cal A}}
\begin{flushright}
OHSTPY-HEP-T-02-006\\
hep-th/0205136
\end{flushright}
\vspace{20mm}
\begin{center}
{\LARGE Scalar propagator in the pp-wave geometry obtained from
$AdS_5\times S^5$}
\\
\vspace{20mm}
{\bf     Samir D. Mathur\footnote{mathur@pacific.mps.ohio-state.edu}, 
Ashish Saxena\footnote{ashish@pacific.mps.ohio-state.edu} and Yogesh 
K. Srivastava\footnote{yogesh@pacific.mps.ohio-state.edu} \\}
\vspace{12mm}
Department of Physics,\\ The Ohio State University,\\ Columbus, OH 43210, USA\\
\vspace{4mm}
\end{center}
\vspace{10mm}
\begin{abstract}

We compute the propagator for   massless and  massive scalar
fields in the metric
of the pp-wave. The retarded
propagator for the massless field is found to stay confined to the
surface formed by  null geodesics.
The algebraic form of the massive propagator is found to be related
in a simple way to the
form of the propagator in flat spacetime.

\end{abstract}
\newpage

\section{Introduction.}

Recently there has been a surge of interest in the study of string
theory on pp-wave backgrounds.
A 10-d pp-wave metric (with a certain background 5-form field
strength) can be obtained as a limit
of $AdS_5\times S^5$: it is the metric seen by particles that rotate
fast around the diameter of the
$S^5$~\cite{blau}.

String states in such backgrounds were studied in \cite{mal}.  To proceed
further and study interactions
between such states one needs to know the propagators of the fields that
are exchanged between string
states. Since we are in Type IIB string theory, the massless bosonic
fields that can be
exchanged are the dilaton $\Phi$ and axion $C$, the NS-NS and RR 2-form gauge
fields $B_{\mu\nu}^{NSNS}$ and
$B^{RR}_{\mu\nu}$, the 4-form gauge field $A_{\mu\nu\lambda\rho}$ and
the graviton
$h_{\mu\nu}$.
In this  letter we compute the propagator for  massless scalar fields
like $\Phi$, as well as for massive scalar fields.

We note that the propagator in $AdS$ space is known \cite{Ostling} ,
and in principle
one can try to extract the
propagator in the pp-wave by starting with the propagators in $AdS$
space for different harmonics
on the sphere, putting together these propagators in an appropriate
way and taking an appropriate
limit. This process appears to be cumbersome though, and we find it
much easier to work directly
with the pp-wave background instead.

Our principal results are the following. We compute the retarded and
Feynman propagators
for the scalar, both for the massless field and the massive field.
The retarded propagator
is found to stay confined to the surface formed by the null geodesics.
   For both the massless and the massive propagators we find
algebraic forms that are
closely related to the forms found
in flat space. The results thus suggest that even for higher spin
fields there might be a simple way to
guess the pp-wave propagators from flat space computations.

One of our goals in finding the exact form of  the propagator was to
study the behavior of highly
excited string states which can collapse into black holes under their
self-gravity.  We comment
briefly on this problem in the discussion. While this paper was in
preparation there appeared \cite{li}
which studied a similar question using an approximate form of the propagator.

\section{Constructing the propagator from solutions of the wave equation}

We consider the 10-D metric obtained from the pp-wave limit of
$AdS_5\times S^5$.
The line element in pp-wave
background is
\begin{equation}
ds^{2}= -4 dx^{+}dx^{-} - r^2 (dx^{+})^{2} +dx_idx_i
\label{five}
\end{equation}
where $x_i, i=1\dots 8$ are the transverse coordinates and $r^2
=x_{i} x_{i}$ is the distance squared in the transverse coordinates.
The scalar laplacian is
\begin{eqnarray}
\nabla^{2} &= &\frac{1}{\sqrt{-g}}\nabla_{a} (g^{ab} \sqrt{-g}
\nabla_{b}) \nonumber \\
&= & \frac{r^{2}}{4} \partial_{-}^{2} - \partial_{-}\partial_{+}
+\nabla^{2}_{8}
\end{eqnarray}
where $\nabla^{2}_{8}=\partial_i\partial_i$.
The  wave-equation for mass $m$ is
\begin{equation}
(\nabla^{2} -m^2)F=0
\end{equation}
To find the propagator we will solve the
eigenvalue problem for the above operator. Let $F_{\lambda}$ be an
eigenfunction with eigenvalue $-\lambda$:
\be
(\nabla^{2} -m^2)F_\lambda=-\lambda F_\lambda
\label{one}
\ee
We write
\be
F_\lambda[x^{+},x^{-}, x_{i}]= \mathcal{N}e^{ik_{+} x^{+} + ik_{-}
x^{-}}f[x_{i}]
\ee
($\cal{N}$ is a normalization) obtaining
\begin{equation}
\nabla_{8}^{2}f -\frac{k_{-}^{2} r^{2}}{4}f + (k_{+}k_{-}- m^{2}+\lambda)f = 0
\end{equation}
This is just the equation for a non-relativistic, spherically symmetric quantum
harmonic oscillator in 8 dimensions
with frequency
\be
\omega=\frac{|k_{-}|}{2}
\label{three}
\ee
      and total energy
\be
E=\frac{(k_{+}k_{-} -
m^{2}+\lambda)}{2}.
\ee
Writing
\be
f=\prod_i f_i
\ee
we get
\begin{equation}
\frac{1}{f_{i}} \frac{d^{2} f_{i}}{dx_{i}^{2}} - \omega^{2} x_{i}^{2} = - g_{i}
\end{equation}
      with
\be
\Sigma_{i=1}^{8} g_{i} = 2E.
\ee
The normalised solutions are
\begin{equation}
f_{i}= \left(\frac{\sqrt{\omega}}{2^{n_{i}} n_{i}!
\sqrt{\pi}}\right)^{\frac{1}{2}}H_{n_{i}}(\tilde{x}_{i})
e^{-\frac{1}{2}\tilde{x}_{i}^{2}}
\end{equation}
      where $\tilde{x}_{i}= \sqrt{\omega}x_{i}$, $H_{n}$ are the Hermite
polynomials of order $n$ and  $g_i=2(n_{i} + \frac{1}{2})\omega$. We can
now write the normalized
solutions to (\ref{one})  as
\begin{equation}
F_\lambda[x^{+},x^{-},x_{i}]= {1\over \sqrt{2}}\frac{1}{2\pi}  e^{ik_{+} x^{+}}
e^{ik_{-} x^{-}} \Pi_{i=1}^{8}
\left(\frac{\sqrt{\omega}}{2^{n_{i}}
n_{i}! \sqrt{\pi}}\right)^{\frac{1}{2}}H_{n_{i}}(\tilde{x}_{i})
e^{-\frac{1}{2}\tilde{x}_{i}^{2}}
\end{equation}
with
\be
\lambda=-[k_{+}k_{-} - |k_{-}|\Sigma_{i=1}^{8} (n_{i}+\frac{1}{2}) -m^{2}]
\ee

      The propagator satisfying
\be
(\triangle -m^2) G(\mathbf{x_{2}}, \mathbf{x_{1}}) = {1\over
\sqrt{-g}}\delta ({\bf
x_2}-{\bf x_1})
\label{thir}
\ee
is then given by ($\sqrt{-g}=2$)
\begin{eqnarray}
G(\mathbf{x_{2}}, \mathbf{x_{1}})
&=&\sum_\lambda {F_\lambda({\bf x_2})F^*_\lambda({\bf x_1})\over (-\lambda)}\cr
      &=&
\int_{-\infty}^{\infty}
\frac{dk_{+} dk_{-}}{2(2\pi)^{2}} e^{ik_{+} (x^{+}_{2}-x^{+}_{1})}
e^{ik_{-}( x^{-}_{2} -x^{-}_{1})}
\nonumber \\ &
&\Sigma_{\{ n_{i}\} }\Pi_{i=1}^{8}
\left(\frac{\sqrt{\omega}}{2^{n_{i}} n_{i}! \sqrt{\pi}}\right)
      \frac{H_{n_{i}}(\tilde{x}_{1i})
e^{-\frac{1}{2}\tilde{x}_{1i}^{2}}H_{n_{i}}(\tilde{x}_{2i})
e^{-\frac{1}{2}\tilde{x}_{2i}^{2}} }{k_{+}k_{-}
- |k_{-}|\Sigma_{i=1}^{8} (n_{i}+\frac{1}{2})-m^{2}}
\end{eqnarray}

\section{Simplifying the expression for the propagator}

Since we encounter the absolute value of $k_-$ in $\omega$ eq.
(\ref{three}),  it is convenient to
break up the $k_-$ integral into two parts: a part $(-\infty,0)$ and
a part $(0, \infty)$.
We write  $n \equiv \Sigma_{i=1}^{8} n_{i}$ and
$\Delta x^{\pm}= x^{\pm}_{2}-x^{\pm}_{1}$. Then

\be
G(\mathbf{x_{2}}, \mathbf{x_{1}})=I_++I_-
\ee
where

\begin{eqnarray}
I_{+}  =  \Sigma_{\{n_{i}\}}{1\over 2}
\frac{1}{(2\pi)^{6}}\int_{0}^{\infty} dk_{-} \frac{k_{-}^{4}
e^{ik_{-}\Delta x^{-}}}{k_{-}}
\int_{-\infty}^{\infty} dk_{+} \frac{e^{ik_{+}
\Delta x^{+}}}{k_{+} - (n+4)-m^{2}/k_{-}} & & \nonumber \\
\Pi_{i=1}^{8} \frac{1}{2^{n_{i}} n_{i}!} H_{n_{i}}(\sqrt{\omega}
x_{1i}) H_{n_{i}}(\sqrt{\omega} x_{2i})
e^{-\frac{\omega}{2}\left(x_{1i}^{2} + x_{2i}^{2}\right)}\label{six}
\end{eqnarray}

and  after a change of variables $k_-\rightarrow -k_-$ we can write  $I_-$ as

\begin{eqnarray}
I_{-}  =  -\Sigma_{\{n_{i}\}} {1\over 2}
\frac{1}{(2\pi)^{6}}\int_{0}^{\infty} dk_{-} \frac{k_{-}^{4}
e^{-ik_{-}\Delta x^{-}}}{k_{-}}
\int_{-\infty}^{\infty} dk_{+} \frac{e^{ik_{+}
\Delta x^{+}}}{k_{+} + (n+4) + m^{2}/k_{-}} & & \nonumber \\
\Pi_{i=1}^{8} \frac{1}{2^{n_{i}} n_{i}!} H_{n_{i}}(\sqrt{\omega}
x_{1i}) H_{n_{i}}(\sqrt{\omega} x_{2i})
e^{-\frac{\omega}{2}\left(x_{1i}^{2} + x_{2i}^{2}\right)}
\end{eqnarray}

Consider any point in the pp-wave spacetime and look at the
light-cone in the infinitesimal vicinity of this
point.  First ignore the $x_i$, and look at the light cone in the 2-D
space $x^+, x^-$. The two lines forming this
cone are $x^+=0$, and $x^-=-\frac{x^{+}}{4x_ix_i}$.  Further, the  vector
$V^\mu$ which  has
$V^+>0$ (and   all other components zero) is timelike in
the metric (\ref{five}); this tells us which of the four sectors
marked out by the two null lines is the
forward light cone.  We find that at least for the infinitesimal
vicinity of the starting point a retarded
propagator can be defined by requiring  that the propagator be
nonzero only for
$\Delta x^+>0$.  We adopt this as our definition of retarded
propagator (we will return to a more complete
discussion of null geodesics later).

Let us begin with $I_+$ and perform the integral
over $k_+$. For $\Delta x^+>0$ we can
close the
$k_+$ contour in the upper half plane, so to get a nonvanishing
result from each pole we shift the
poles to slightly above the real axis.  The
$k_+$ integral  just sets
$k_+=(n+4)+m^2/k_-$.

Let us define
\be
z=e^{i\Delta x^+}
\ee
For any one choice of the index $i$ the sum in $\Sigma_{\{n_{i}\}}$
over the Hermite polynomials can
be evaluated by using the  identity \cite{bateman}
\begin{equation}
\Sigma_{n_{i}=0}^{\infty} \frac{H_{n_{i}}(\sqrt{\omega}x_{1i})H_{n_{i}}
(\sqrt{\omega}x_{2i}) (\frac{z}{2})^{n_{i}}}{n_{i}!}= \frac{1}{\sqrt{1-z^{2}}}
e^{\omega\left(\frac{2x_{1i}x_{2i}z -x_{1i}^{2}z^{2}
-x_{2i}^{2}z^{2}}{1-z^{2}}\right)}
\label{eight}
\end{equation}

We need to take a product over 8 values of the index $i$. We then find

\begin{equation}
I_{+}= \frac{2\pi i}{(2\pi)^{6}}
\frac{z^{4}}{(1-z^{2})^{4}}\frac{\Theta(\Delta x^{+})}{2}
\int_{0}^{\infty}dk_{-}  k_{-}^{3} e^{-ik_{-}
Y^{2}}e^{i\frac{m^{2}\Delta x^{+}}{k_{-}}} \label{ten}
\end{equation}

where
\begin{displaymath}
Y^{2}= - \Delta x^{-} + \frac{x_{1}^{2}+x^{2}_{2}}{4i} +
\frac{x_{1}^{2} z^{2} + x_{2}^{2} z^{2} -2 x_{1}. x_{2}
z}{2i(1-z^{2})}
\end{displaymath}

Before proceeding further with the evaluation of   $I_+$ we take note
of the regularizations required
to define $I_+$.  It is a general feature of Green's functions that
for sufficiently high spacetime
dimensionality the Green's function is not square integrable, and so
the Fourier transform need not
be given by a convergent integral --  a damping factor must be
introduced to cut off high
frequencies. Lifting this cutoff at the end  brings the Green's
function back to its required form as
a singular `distribution'.

The damping at high frequencies can be provided by introducing into
$I_+, I_-$ the factors:
\be
      e^{-\epsilon |k_-|}e^{-\epsilon' n_i}
\label{nine}
\ee
The first  of these factors leads to the change in (\ref{ten})
\be
Y^2\rightarrow Y^2-i\epsilon
\ee

The second  factor in (\ref{nine})  leads to the change that in the identity
(\ref{eight}) we have the replacement
\be
z\rightarrow ze^{-\epsilon'}
\ee

With these regulations we find
\begin{equation}
I_{+}= \frac{\Theta(\Delta x^{+})}{2}\frac{2\pi i}{(2\pi)^{6}}\frac{1}{(2\Sin\,
(\Delta x^{+}+i\epsilon'))^{4}}
\int_{0}^{\infty} dk_{-} \ k_{-}^{3} e^{-i(k_{-}(Y^{2}-i\epsilon)
-\frac{m^{2}\Delta x^{+}}{k_{-}})}
\label{Iplus}
\end{equation}
and
\begin{equation}
      Y^{2}=\frac{\Phi}{4\Sin\,(\Delta x^{+}+i\epsilon)}
\end{equation}

where
\begin{equation}
\Phi= -4\Delta x^{-}\Sin \,(\Delta x^{+}+i\epsilon')
-2\Sin^{2}\frac{(\Delta x^{+}+i\epsilon')}{2} (x_{1}^{2}
     +x_{2}^{2}) +(x_{1} -x_{2})^{2}
\end{equation}
is an expression that goes over to the invariant distance squared for
small separations.

Proceeding similarly for $I_-$, we require again that the
contribution be nonzero only for $\Delta
x^+>0$, and we get

\begin{equation}
I_{-}= -\frac{\Theta(\Delta x^{+})}{2}\frac{2\pi
i}{(2\pi)^{6}}\frac{1}{(2\Sin\,
(\Delta x^{+}-i\epsilon'))^{4}}
\int_{0}^{\infty} dk_{-} \ k_{-}^{3} e^{i(k_{-}(Y^{2}+i\epsilon)
-\frac{m^{2}\Delta x^{+}}{k_{-}})}
\label{Iminus}
\end{equation}

Note that  $I_{-}$ is precisely the complex
conjugate of $I_+$.

\section{Closed form expressions for the propagators}

\subsection{The massless propagator}

Let us first work out the case $m^2=0$.  From (\ref{Iplus}) we find that
\begin{equation}
I_{+}= \frac{\Theta(\Delta x^{+})}{2}\frac{2\pi i}{(2\pi)^{6}}\frac{1}{(2\Sin\,
(\Delta x^{+}+i\epsilon'))^{4}}
{6\over (Y^2-i\epsilon)^4}
=
3 \Theta(\Delta x^{+})\frac{2\pi
i}{(2\pi)^{6}} \frac{2^{4}}{(\Phi
-i\tilde\epsilon)^{4}}
\end{equation}
where
\be
\tilde\epsilon=(4\sin\Delta x^+)\epsilon
\ee

Similarily,
\be
I_++I_-=
3 \Theta(\Delta x^{+})\frac{2\pi
i}{(2\pi)^{6}}\left[ \frac{2^{4}}{(\Phi-i\tilde\epsilon)^{4}}
-\frac{2^{4}}{(\Phi
+i\tilde\epsilon)^{4}}\right]
\ee

Noting that
\begin{equation}
\frac{1}{(\Phi-i\tilde\epsilon)}-\frac{1}{(\Phi +i\tilde\epsilon)} =
2\pi i ~({\rm sign}~ \tilde\epsilon)\delta(\Phi)
\end{equation}
we get
\begin{equation}
G_{ret}({\bf x_2}, {\bf x_1}) = \Theta(\Delta
x^{+})\frac{1}{2\pi^{4}}~{\sin\Delta x^+\over |\sin\Delta
x^+|}~\delta^{'''}(\Phi)
\label{seven}
\end{equation}
where the derivatives of the delta function are with respect to its
argument $\Phi$.

It is readily verified that this propagator has the correct
normalization to agree at short
distances with the retarded propagator satisfying (\ref{thir})  (with $m^2=0$).

\subsection{The massive propagator}

Let us now consider the expression (\ref{ten}) for $I_+$ for the
massive scalar field.  Performing the
integral we find \cite{GR}
\begin{equation}
I_{+}= -\frac{\pi^{2}}{16}\Theta(\Delta
x^{+})\frac{1}{(2\pi)^{6}}\frac{ m^{4}(\Delta
x^{+})^{2}}{(Y^{2}-i\eps)^{2}\Sin^{4} (\Delta x^{+}+i\eps') }
     \  H^{(1)}_{-4}(2 m \sqrt{-(Y^{2} -i\epsilon)}\sqrt{\Delta x^{+}})
\label{tw}
\end{equation}
where $H^{(1)}$ is the Hankel function of the first kind. For $I_-$
we get the complex conjugate of $I_+$

\begin{equation}
I_{-}= \frac{\pi^{2}}{16}\Theta(\Delta
x^{+})\frac{1}{(2\pi)^{6}}\frac{ m^{4}(\Delta
x^{+})^{2}}{(Y^{2}+i\eps)^{2}\Sin^{4} (\Delta x^{+}-i\eps') }
       \  H^{(1)}_{-4}(-2 m \sqrt{-(Y^{2} +i\epsilon)}\sqrt{\Delta
x^{+}})\label{tw1}
\end{equation}

Note that the Hankel function is $H^{(1)}_{-4}=J_{-4}+iN_{-4}$. The
function $N_{-4}(z)$ has a
logarithmic branch cut at $z=0$, so in principle we must decide which
branch of the Hankel function
we are on. With the given regulations in
      (\ref{ten}) the integral is well defined for all values of $\Delta
x^+>0, m^2>0$, which suggests that
there should be no ambiguity in the choice of branch, and that we
should be on the principal branch
which is defined by the integral.

To observe that this is indeed what happens, we  look at the behavior
of the argument of the
Hankel function  when
$\Delta x^+$ passes through
$n\pi$ (which causes
$\sin
\Delta x^+$ to vanish) or when we enter or leave the light cone.  We
find that with the given
regulations the argument of the Hankel function  in
$I_+$  stays (for all values of the coordinates) in the
first quadrant of the complex plane.
Since the argument does not thus circle the origin, we stay on the
same branch of the Hankel
function. This branch is given by continuing
the value of $H^{(1)}$  for real positive arguments to the first
quadrant of the complex plane.
A similar analysis holds for $I_-$, and so no ambiguity exists in the
value of the retarded Green's
function which is
\be
G_{ret}=I_++I_-
\ee
with $I_+, I_-$ given by (\ref{tw}), (\ref{tw1}).

We observe that as expected the choice of branch indicated by the
regularizations does not depend on the ratio of the
two regularizations $\epsilon, \epsilon'$, and from now on we just
write $\epsilon$ for both.

\subsection{The Feynman propagator}

To define the analogue of the flat space Feynman propagator we
require that positive frequencies in
the variable $x^+$ travel forward in $x^+$ and negative frequencies
travel backwards in $x^+$.
To make $I_+$ nonzero at negative $\Delta x^+$ we must shift the
poles in $k_+$ integral in (\ref{six})
into the lower half plane. We then get instead of (\ref{tw}) the result
\begin{equation}
\tilde{I_{+}}= \frac{\pi^{2}}{16}\Theta(-\Delta
x^{+})\frac{1}{(2\pi)^{6}}\frac{ m^{4}(\Delta
x^{+})^{2}}{(Y^{2}-i\eps)^{2}\Sin^{4} (\Delta x^{+}+i\eps') }
     \  H^{(1)}_{-4}(2 m \sqrt{-(Y^{2} -i\epsilon)}\sqrt{\Delta x^{+}})
\end{equation}
The Feynman propagator is then
\be
G_F({\bf x_2}, {\bf x_1})=\tilde I_++I_-
\ee

For completeness we record the result of shifting poles in $I_-$ such
that we get a contribution only
for $\Delta x^+<0$:

\begin{equation}
\tilde{I_{-}}= -\frac{\pi^{2}}{16}\Theta(-\Delta
x^{+})\frac{1}{(2\pi)^{6}}\frac{ m^{4}(\Delta
x^{+})^{2}}{(Y^{2}+i\eps)^{2}\Sin^{4}(\Delta x^{+}-i\epsilon)}\
H^{(1)}_{-4}(-2 m
\sqrt{-(Y^{2}+i\epsilon)}\sqrt{\Delta x^{+}})
\end{equation}

\section{Similarities with  flat space propagators}

\subsection{The massless propagator}

If spacetime is flat and its total dimension is even then the
retarded  propagator $G({\bf x_2}, {\bf
x_1})$ for massless fields  stays confined to the null cone, which is
the cone generated by
the tracks of null geodesics starting at ${\bf x_1}$.  We will now
observe that a similar property holds
for the retarded massless propagator in the pp-wave.

Let us first compute the geodesics in the pp-wave metric. If $\tau$ is
an affine parameter along the
geodesic, then the geodesic equations are
\be
{d^2 x^+\over d\tau^2}=0
\ee
\begin{equation}
\frac{d^{2} x^{-}}{d\tau^{2}} + \dot x^+\sum_{i=1}^{8} x_{i} \dot{x_{i}}= 0
\end{equation}
\begin{equation}
\frac{d^{2} x_{i}}{d\tau^{2}} + (\dot x^+)^2x_{i}= 0
\end{equation}
where a dot denotes the derivative with respect to $\tau$.
The first equation gives
      \be
x^+=\alpha\tau+\beta
\label{el}
\ee

If $\alpha\ne 0$, then we can scale $\tau$ to set
$x^+=\tau$.
(For massive geodesics $\tau$ will not be the proper distance
along the geodesic but rather a multiple of the proper distance.)
For the initial conditions
\begin{eqnarray}
x^{-}(\tau=0) & = & x^{-}_{1} \nonumber \\
\frac{d x^{-}}{d\tau}(\tau=0) &=& \dot{x}^{-}_{1} \nonumber \\
x_{i}(\tau=0) &=& x_{1i} \nonumber \\
\frac{d x_{i}}{d\tau} &= & \dot{x}_{1i}
\end{eqnarray}
we get the solution
\begin{eqnarray}
x^{-}(\tau) - x^{-}_{1}&= &
\frac{\dot{\vec{x}_{1}}^{2}-\vec{x_{1}}^{2}}{8} \Sin \, 2\tau +
      \frac{\dot{\vec{x}_{1}}.x_{1}}{4} \Cos\,2\tau +C\tau-
\frac{\dot{\vec{x}}_{1}.x_{1}}{4} \\
\vec{x}(\tau) &=& \dot{\vec{x}_{1}} \,\Sin\, \tau + \vec{x_{1}}\, \Cos\,\tau
\end{eqnarray}
where $\vec{x}$ denotes the coordinates in the  transverse $8$
dimensional space, and
\be
C=\dot{x}^{-}_{1}-\frac{\dot{\vec{x}}_{1}^{2} - \vec{x}_{1}^{2}}{4}
\ee
For null geodesics the initial conditions imply that $C=0$. A little
algebra then shows that
along such geodesics
\be
\Phi({\bf x}, {\bf x_1})=0
\ee
so that the retarded propagator (\ref{seven}) is confined to the
surface formed by the null
geodesics.\footnote{If $\alpha=0$ in (\ref{el}) then we have the solution
$
x^+=x^+_1,
x^-= \gamma\tau+\delta,
{\vec x}={\vec A} \tau +{\vec B}
$.
These geodesics cannot be timelike, and the only null geodesic is
(after scaling $\tau$)
$x^+=const., ~x^-=\tau, ~ {\vec x}={\vec x_1}$. This geodesic is a
limiting case of those obtained for
$\alpha\ne 0$, and so does not affect the conclusion above regarding
the surface formed by the
null geodesics.}

In flat space the propagator is a function of the
invariant distance squared
\be
s^2_{flat}=-4 \Delta x^-\Delta x^+ + (\vec x_1-\vec x_2)^2.
\ee

We have
\begin{equation}
G^{flat}_{ret}({\bf x_2}, {\bf x_1}) =
\Theta(t)\frac{1}{2(\pi)^{4}}\delta^{'''}(-s^2)
\end{equation}

We thus see that the pp-wave result (\ref{seven}) is related to the
flat space result  by the
replacement
$(-s^2_{flat})\rightarrow\Phi$.

\subsection{The massive propagator}

For $\Delta x^+<<1$ we can replace $\sin \Delta x^+\rightarrow \Delta
x^+$ and the  propagator reduces to the
propagator in flat space, as expected.
     What is
interesting  is that the algebraic form of the massive propagator in
the pp-wave is closely related to the
form of the propagator in flat space.

For flat space
\be
I_+^{flat}= -\frac{\pi^{2}}{16}\Theta(\Delta
x^{+})\frac{1}{(2\pi)^{6}}\frac{ 16 m^{4}}{(-s^2_{flat})^2}
     \  H^{(1)}_{-4}(m \sqrt{-s^2_{flat}})
\label{fourt}
\ee
with similar expressions  for the other functions involved in the propagators.
Now note that for a timelike geodesic
\be
{ds^2\over (dx^+)^2}=-4C
\ee
where the constant $C$ can be written in terms of the initial and
final points as
\be
C=-{\Phi\over 4 \Delta x^+\sin\Delta x^+}
\ee
Thus the distance
   measured along a timelike geodesic will  be
\be
     \int_{\bf x_1}^{\bf x_2}  \sqrt{-ds^2}=\sqrt{-\Phi {\Delta
x^+\over \sin\Delta
x^+}}=2\sqrt{-Y^2}\sqrt{\Delta x^+}\equiv \sqrt{-\sigma^2}
\ee

We observe that the argument of the Hankel function in  $I_+$ is $ m
{\bf \sqrt{-\sigma^2}}$. Overall we can write
\be
I_+= -\frac{\pi^{2}}{16}\Theta(\Delta
x^{+})\frac{1}{(2\pi)^{6}}\left({\sin \Delta x^+\over \Delta x^+}\right)^{-4}\frac{
16 m^{4}}{(-\sigma^2)^2}
     \  H^{(1)}_{-4}(m \sqrt{-\sigma^2})
\ee
which differs from  (\ref{fourt}) only by the factor
$({\sin \Delta x^+\over \Delta x^+})^4$ and the replacement 
$s_{flat}^2\rightarrow
\sigma^2$.

Let us now observe the relation to the flat space propagator in
another way. In the massless case we had found that the
relation to the flat space propagator was directly seen when using the variable
$\Phi$.  Define
\be
\tilde m^2 = m^2 {\Delta x^+\over \sin\Delta x^+}
\ee
Then we observe that
\be
I_+= -\frac{\pi^{2}}{16}\Theta(\Delta
x^{+})\frac{1}{(2\pi)^{6}}\frac{ 16 \tilde m^{4}}{(-\Phi)^2}
     \  H^{(1)}_{-4}( \tilde m \sqrt{-\Phi})
\ee
which differs from (\ref{fourt}) only by the replacements
$m\rightarrow \tilde m$ and $ s^2_{flat}\rightarrow \Phi$.

\section{Discussion}

We have obtained propagators for the massless and massive scalar
fields in the pp-wave geometry obtained from $AdS_5\times S^5$.
This is the geometry seen by string states moving fast around the
diameter of the $S^5$, while staying in the vicinity of the origin in
global
$AdS_5$. We have noted several relations between these propagators
and the corresponding flat space expressions. These relations should
help us
to guess the form of propagators for higher spin fields.  The
propagators for other spaces $AdS_p\times S^q$ can be obtained by
direct inspection of
our results here -- the different dimension leads to a change in  the
number of harmonic oscillators obtained from the transverse
coordinates
$x_i$, but there is no other essential change in the computation.

The space $AdS_5\times S^5$ is locally
conformally flat, so for massless fields one should be able to
recover the propagator in the pp-wave from the flat space
result.\footnote{We thank D.
Berenstein for this comment.} (We have not taken a conformally
coupled scalar, but the curvature scalar $R$ of the geometry vanishes
so we do not
have to consider a term $R\phi^2$  in the Lagrangian for a scalar
field $\phi$.)  Working with such conformal maps may make it more difficult
however to follow the behavior of the propagator through the singularities at
$\Delta x^{+} = n\pi$, since the conformal map is local rather than global.
The massive propagator has a length scale set by the mass, and so  cannot be
obtained by a conformal mapping to flat space.

The pp-wave has the property that light rays can reach the boundary
and return in a finite time, a property that is also shared by $AdS$
spaces.
We thus need to have a boundary condition at infinity which reflects
the energy of the waves back into the geometry. The fact that we have
used
normalizable wavefunctions in constructing the propagator implies that
we have Dirichlet boundary conditions at infinity.

One of the interesting features of the propagators is the appearance
of the function $\sin \Delta x^+$, which  vanishes whenever
$\Delta x^+=n\pi$.
Let us write
\be x^+={1\over 2} (t+\psi), ~~x^-={1\over 2}(t-\psi)
\ee
The pp-wave geometry (\ref{five}) is infinitely extended in the
coordinate $\psi$ along the wave. However when we consider this
geometry as a part
of $AdS_5\times S^5$ then we find that $\psi$ is an angular variable
on a circle. The point $x^+=\pi, x^-=x_i=0$ is  at $\psi=\pi$,
halfway around the
$S^5$. The point $x^+=2\pi, x^-=x_i=0$ is the same as the point
$x^+=0, x^-=x_i=0$. Null geodesics moving on the $S^5$ cross the line
$x_i=0$ twice in
each revolution, at the points $x^+=0, x^+=\pi$.

Finally we comment on the physical problem that we would like to
address with the help of this propagator (and similar propagators for
other fields).
In \cite{mal} string states were considered that move fast around the
$S^5$.  If we consider higher and higher excitations of this string,
then there
can come a point where the string collapses to a black hole under its
self-gravitation. Such an effect was studied for strings in flat
space in \cite{hp}.
Let us note here some aspects of this problem in the pp-wave geometry.

In flat space the size of the free string is given by a `random walk'
made out of string bits. In the pp-wave geometry there is a
potential that drives
the string to the center $x_i=0$. There is no potential confining the
direction $\psi$ along the wave, so the string can extend into a
cylindrical  shaped
object lying along the central axis of the pp-wave.

Let us assume such a geometry for the string state, and consider the
effects of self-interaction created by a field with a propagator that
behaves like
that of the massless scalar field considered here. (We expect the
dilaton, 2-form gauge field an the graviton to have propagators that
are similar
except for their index structure.)  Note that the massless propagator
is a function of the variable $\Phi$, and  consider the propagator
between
points that are near the axis
$x_i=0$. The fact that $\Phi\sim -4 \Delta x^-\sin \Delta x^+$ implies
that the retarded Green's function behaves like an inverse power of
$\sin x^+$ instead of an inverse power of $x^+$.

Now consider applying such a propagator to find the self-interaction
of a time independent distribution. We find a `resonant interaction'
between
points that are separated by  $\Delta \psi=n\pi$.  Such an
interaction permits solutions that are independent of the coordinate
$\psi$, but it also
suggests that for suitable interactions one may get lower energy
configurations that are periodic under $\psi\rightarrow \psi+n\pi$
rather than
independent of $\psi$.  We hope to return to a more detailed
discussion of this problem  elsewhere.

\section*{Acknowledgments}

We are grateful  to David Berenstein,  Camillo Imbimbo, Oleg Lunin
and Amit K. Sanyal for several
helpful comments. This work is supported in part by DOE
grant DE-FG02-91ER40690.

\end{document}